\documentclass{iaus}
\usepackage[dvips]{graphicx}
  \checkfont{eurm10}
  \iffontfound
    \IfFileExists{upmath.sty}
      {\typeout{^^JFound AMS Euler Roman fonts on the system,
                   using the 'upmath' package.^^J}%
       \usepackage{upmath}}
      {\typeout{^^JFound AMS Euler Roman fonts on the system, but you
                   dont seem to have the}%
       \typeout{'upmath' package installed. iaus.cls can take advantage
                 of these fonts,^^Jif you use 'upmath' package.^^J}%
      }
  \else
  \fi

  \checkfont{msam10}
  \iffontfound
    \IfFileExists{amssymb.sty}
      {\typeout{^^JFound AMS Symbol fonts on the system, using the
                'amssymb' package.^^J}%
       \usepackage{amssymb}%
         \let\leq=\leqslant
       \let\ge=\geqslant  \let\geq=\geqslant
      }{}
  \fi

  \IfFileExists{amsbsy.sty}
    {\typeout{^^JFound the 'amsbsy' package on the system, using it.^^J}%
     \usepackage{amsbsy}}
    {}

\newsavebox{\astrutbox}
\sbox{\astrutbox}{\rule[-5pt]{0pt}{20pt}}

\title[Dominant galaxies in 2dF groups]
	{Dominant galaxies in 2dF groups}
\author[B. Kelm et al.]%
{B. Kelm %
\and P. Focardi}
\affiliation{Dipartimento di Astronomia, Universit\'a di Bologna }
 \pubyear{2004}
\volume{195}
\pagerange{1--8}
\date{?? and in revised form ??}
\jname{Outskirts of Galaxy Clusters: intense life in the suburbs}
\editors{A. Diaferio, ed.}
\begin{document}
\maketitle
\begin{abstract}
We investigate whether the spectral-type of a locally dominant 
(most luminous) galaxy can be used to select sets of galaxies that are 
physically associated (groups).  
We assume that passive dominants trace a group-like potential, 
and SF-dominants a field-like environment.  
The group sample includes 988 groups 
selected in the 2dFGRS applying a maximum magnitude difference criterion. 
We find that the average number of passive galaxies associated 
to a dominant is larger when the dominant is passive,    
a result supporting our assumption that galaxy associations 
around a passive dominant are reliable groups.  
Finally we show that, to reduce the contamination by unbound galaxy 
associations (SF-dominant), a $\geq$3 passive-members criterion is more 
efficient than a $\geq$6 all-members criterion.   
\end{abstract}
\section{Dominant Group Galaxy selection criteria}
Each 2dF galaxy with $\geq$4 neighbours (projected separation less than 
1$h^{-1}$Mpc and line of sight velocity difference less than 
1000\,km\,s$^{-1}$), that is more luminous (by $\geq$0.2 mag) than 
all of its neighbours, is a dominant group galaxy and identifies a group.  
Given the bright and faint apparent magnitude limits of the 2dF, 
a maximum magnitude difference criterion has been applied providing, for 
each dominant, complete identification of all neighbours from 
$\sim$equally-luminous to $\sim$2$\div$2.5 magnitudes fainter. 
Dominants are selected in the range $b_{j}$$\in$[17 $\div$ 17.5], 
 neighbours in range $b_{j}$$\in$[17 $\div$ 19.5]. 
The criteria for selection are 
0.03 $\leq$ $z_{dom}$ $\leq$ 0.12 and $-22\leq (M - 5 log h)_{dom} 
\leq-19$. 
The sample includes 988 groups (7281 galaxies): 639 with a passive 
(Type 1 in Madgwick et al. 2002) dominant galaxy (P-dG), and 349 with a 
SF (Type$>$1) dominant (SF-dG). P-dGs are typically richer than SF-dGs.      
\section{Passive and Star-Forming dominants}
If we assume that P-dGs trace a group-like potential, and SF-dGs a 
field-like environment (galaxy associations not embedded within a common 
massive halo), differences in the spectral mix of P-dGs and SF-dGs members 
should reflect the role of a group environment on the spectral evolution 
of galaxies. 
\begin{figure}
{\includegraphics[width=2.5in]{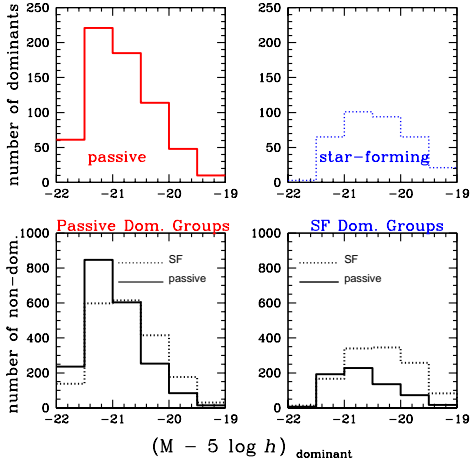}} 
\caption{Histograms show the abundance of group dominants 
({\it top panels}) and group non-dominants ({\it bottom panels}) as a 
function of the absolute magnitude of the dominant.}
\end{figure}
Figure 1 shows the abundance of P-dGs, and SF-dGs 
({\it top panels}) as a function of absolute magnitude  of the dominant. 
Non-luminous dominants are equally likely to be passive or SF, 
but the former are much more common than the latter among luminous dominants.  
That SF galaxies represent only a small fraction among 
luminous dominant galaxies is consistent with X-ray studies that show  
diffuse hot X-ray halo typically occur in groups hosting a dominant 
luminous elliptical. It remains unclear whether groups with a 
faint passive dominant exist at all, or whether these are typically 
embedded within larger systems (\cite{Kelm04}).    

The {\it bottom panels} show the distributions of non-dominants that are 
passive and star-forming, as a function of the luminosity of the dominant, 
in P-dGs ({\it left}) and SF-dGs ({\it right}). 
P-dGs with a luminous dominant exhibit more passive 
than SF non-dominant members, but no passive-member excess is seen 
in P-dGs with a faint (M$\ge$-20.5) dominant. 
Conversely, SF-dGs exhibit more non-dominant SF members than 
passive ones, except for very luminous (M$\leq-21$) dominants. 
Being well established that SF activity decreases with increasing local 
density (\cite{Lewis02,Gomez03}) the observed trends in figure 1 might be a 
further hint that, when luminous, P-dGs are more massive systems than SF-dGs. 

In figure 2 (top panels) the average fractional content in passive 
({\it left}) and SF ({\it right}) galaxies is plotted as a function of the 
dominant luminosity in P-dGs ({\it solid}) and SF-dGs ({\it dotted}). 
SF galaxies are less frequent in P-dGs, a result  
that supports our assumption that only passive dominants are key 
tracer of group-like potentials. 
It further agrees with recent result (\cite{Balogh04}) that a 
lower level of SF activity in high density environments is largely due to 
the smaller fraction of SF galaxies in these environments. 
However, the fractions of passive and SF galaxies in groups definitely 
appear to depend on the luminosity of the dominant, rather than on its 
spectral type. The modest difference observed in the SF-galaxy fraction 
between P-dGs and SF-dGs suggests that, in group samples,  
the SF-fraction is generally less efficient than the dominant-spectral-type 
criterion to segregate a group-like halo from galaxy-size halo associations. 
\section{Passive dominated groups: is SF depressed?}
The slightly lower SF fraction associated to P-dGs might derive from 
a decrease in the number of SF galaxies or from an increase in 
the number of passive galaxies. In figure 2 we plot ({\it bottom panels}) 
the average number of passive ({\it left}) and SF ({\it right}) non-dominants 
associated to each dominant as a function of the dominant luminosity.  
At all luminosities, P-dGs exhibit no deficit of SF non-dominants relative 
to SF-dGs.  Conversely, 
the average number of passive non-dominants per dominant is larger 
in P-dGs than in SF-dGs (figure 2 {\it bottom left}). This indicates that  
 the clustering of passive non-dominants is larger in P-dGs than in SF-dGs.  
Results are clearly consistent with the hypothesis that only P-dGs are 
tracer of group-like potentials. 
The excess of passive galaxies in P-dGs further supports a biased galaxy 
formation scenario %\cite{Sommerville99,Kodama04,Sheth04}, 
in which the formation of massive galaxies (mass assembly and 
star-formation), from the highest peaks in the initial density fluctuation 
field, is predicted to occur in an accelerated way (or at earlier time) in 
systems embedded within larger halos. 
Projection effects could be responsible for the similar number of non-dominant 
SF galaxies in the field and in groups. Alternatively, substantial infall 
on P-dGs from field  SF-galaxies is required, thereby 
matching the prediction that the fraction of mass 
in the universe bound in group-like systems has undergone a 
dramatic increase between z=1 and z=0.  

Figure 2 ({\it bottom panels}) also provides evidence that the 
average number of SF galaxies is less sensible than the average number 
of passive galaxies, to the luminosity of the dominant 
(--$>$ mass of the system). 
This confirms that the relation between density and luminosity is 
weaker in SF than in passive galaxies (\cite{Hogg03,Balogh04}).  
%%%%%%%%%%%%%%%%%%%%%%%%%%%%%%%%%%%%%%%%%%%%%%%%%%%%%%%%%
\section{Group selection criteria: how to find less SF-dominant groups } 
We have shown that SF-dGs are generally unfair tracer of group-like 
potentials.  
Three different selection criteria have been applied to the non-dominant 
members in the 988 groups to explore which one more efficiently reduces the 
number of SF-dominant groups: 

\noindent i) {\bf $\ge$6 non-dom} $->$ 471 groups: 330 P-dGs and 141 
SF-dGs %(P$_{dom}$/SF$_{dom}$=2.3)

\noindent ii) {\bf $\ge$3 passive non-dom} $->$ 426 groups: 330 P-dGs 
and 96 SF-dGs %(P$_{dom}§$/SF$_{dom}$=3.4)

\noindent iii) {\bf $\ge$50\% passive non-dom} $->$ 455 groups: 337 
P-dGs and 118 SF-dGs %(P$_{dom}$/SF$_{dom}$=2.9)

Clearly, the request for {\bf $\ge$3 passive non-dom} is the most 
efficient criterion in rejecting groups with a SF-dominant. It is also less 
biased by projections of field-galaxies and it additionally produces groups 
with a passive galaxy fraction as high as 50\%. 
\begin{figure}
{\includegraphics[width=2.5in]{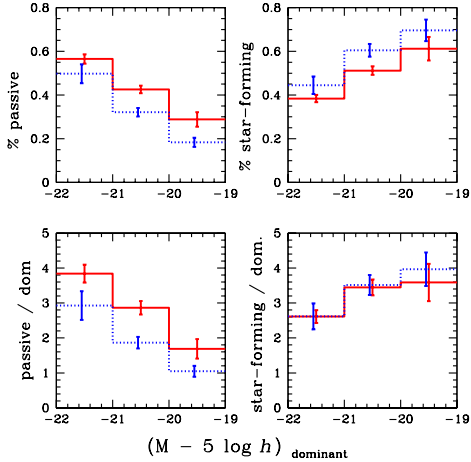}} 
\caption{Fractions of passive and SF non-dominants ({\it top panels}) among 
non-dominants in groups with a passive ({\it solid}) and 
a SF ({\it dotted}) dominant. The {\it bottom panels} show the average 
number of passive and SF non-dominants associated to each dominant. 
Error bars are multinomial.}
\end{figure}

\end{document}